\PassOptionsToPackage{unicode}{hyperref}
\PassOptionsToPackage{hyphens}{url}
\PassOptionsToPackage{dvipsnames,svgnames,x11names}{xcolor}
\documentclass[
  12pt]{article}

\usepackage{amsmath,amssymb}
\usepackage{iftex}
\ifPDFTeX
  \usepackage[T1]{fontenc}
  \usepackage[utf8]{inputenc}
  \usepackage{textcomp} % provide euro and other symbols
\else % if luatex or xetex
  \usepackage{unicode-math}
  \defaultfontfeatures{Scale=MatchLowercase}
  \defaultfontfeatures[\rmfamily]{Ligatures=TeX,Scale=1}
\fi
\usepackage{lmodern}
\usepackage{authblk}
\ifPDFTeX\else  
\fi
\IfFileExists{upquote.sty}{\usepackage{upquote}}{}
\IfFileExists{microtype.sty}{% use microtype if available
  \usepackage[]{microtype}
  \UseMicrotypeSet[protrusion]{basicmath} % disable protrusion for tt fonts
}{}
\makeatletter
\@ifundefined{KOMAClassName}{% if non-KOMA class
  \IfFileExists{parskip.sty}{%
    \usepackage{parskip}
  }{% else
    \setlength{\parindent}{0pt}
    \setlength{\parskip}{6pt plus 2pt minus 1pt}}
}{% if KOMA class
  \KOMAoptions{parskip=half}}
\makeatother
\usepackage{xcolor}
\makeatletter
\ifx\paragraph\undefined\else
  \let\oldparagraph\paragraph
  \renewcommand{\paragraph}{
    \@ifstar
      \xxxParagraphStar
      \xxxParagraphNoStar
  }
  \newcommand{\xxxParagraphStar}[1]{\oldparagraph*{#1}\mbox{}}
  \newcommand{\xxxParagraphNoStar}[1]{\oldparagraph{#1}\mbox{}}
\fi
\ifx\subparagraph\undefined\else
  \let\oldsubparagraph\subparagraph
  \renewcommand{\subparagraph}{
    \@ifstar
      \xxxSubParagraphStar
      \xxxSubParagraphNoStar
  }
  \newcommand{\xxxSubParagraphStar}[1]{\oldsubparagraph*{#1}\mbox{}}
  \newcommand{\xxxSubParagraphNoStar}[1]{\oldsubparagraph{#1}\mbox{}}
\fi
\makeatother

\usepackage{longtable,booktabs,array}
\usepackage{calc} % for calculating minipage widths
\usepackage{etoolbox}
\makeatletter
\patchcmd\longtable{\par}{\if@noskipsec\mbox{}\fi\par}{}{}
\makeatother
\IfFileExists{footnotehyper.sty}{\usepackage{footnotehyper}}{\usepackage{footnote}}
\makesavenoteenv{longtable}
\usepackage{graphicx}
\makeatletter
\def\maxwidth{\ifdim\Gin@nat@width>\linewidth\linewidth\else\Gin@nat@width\fi}
\def\maxheight{\ifdim\Gin@nat@height>\textheight\textheight\else\Gin@nat@height\fi}
\makeatother
\setkeys{Gin}{width=\maxwidth,height=\maxheight,keepaspectratio}
\makeatletter
\def\fps@figure{htbp}
\makeatother

\makeatletter
\@ifpackageloaded{caption}{}{\usepackage{caption}}
\AtBeginDocument{%
\ifdefined\contentsname
  \renewcommand*\contentsname{Table of contents}
\else
  \newcommand\contentsname{Table of contents}
\fi
\ifdefined\listfigurename
  \renewcommand*\listfigurename{List of Figures}
\else
  \newcommand\listfigurename{List of Figures}
\fi
\ifdefined\listtablename
  \renewcommand*\listtablename{List of Tables}
\else
  \newcommand\listtablename{List of Tables}
\fi
\ifdefined\figurename
  \renewcommand*\figurename{Figure}
\else
  \newcommand\figurename{Figure}
\fi
\ifdefined\tablename
  \renewcommand*\tablename{Table}
\else
  \newcommand\tablename{Table}
\fi
}
\@ifpackageloaded{float}{}{\usepackage{float}}
\floatstyle{ruled}
\@ifundefined{c@chapter}{\newfloat{codelisting}{h}{lop}}{\newfloat{codelisting}{h}{lop}[chapter]}
\floatname{codelisting}{Listing}

\makeatother
\makeatletter
\@ifpackageloaded{caption}{}{\usepackage{caption}}
\@ifpackageloaded{subcaption}{}{\usepackage{subcaption}}
\makeatother

\usepackage{amsthm}
\usepackage{makecell}

\usepackage{booktabs}      % \toprule, \midrule, \bottomrule
\usepackage{threeparttable} % table footnotes
\usepackage{booktabs}
\usepackage[most]{tcolorbox}
\usepackage{xcolor}

\newtheorem{assumption}{Assumption}

\newcommand{\T}{\top}

\usepackage{xcolor}
\usepackage{listings}

\lstdefinestyle{rcode}{
  language=R,
  basicstyle=\ttfamily\small,
  commentstyle=\color{green!50!black},
  stringstyle=\color{red},
  breaklines=true,
  columns=fullflexible,
  frame=single,
  backgroundcolor=\color{gray!8},
  showstringspaces=false
}

\lstnewenvironment{rcode}{\lstset{style=rcode}}{}

\ifLuaTeX
  \usepackage{selnolig}  % disable illegal ligatures
\fi
\usepackage[]{natbib}
\setcitestyle{authoryear,round}
\usepackage{bookmark}

\IfFileExists{xurl.sty}{\usepackage{xurl}}{} % add URL line breaks if available
\hypersetup{
  pdftitle={Title},
  pdfauthor={Author 1; Author 2},
  pdfkeywords={3 to 6 keywords, that do not appear in the title},
  colorlinks=true,
  linkcolor={blue},
  filecolor={Maroon},
  citecolor={Blue},
  urlcolor={Blue},
  pdfcreator={LaTeX via pandoc}}

\usepackage{silence}
\newcommand{\anon}{1}

\begin{document}

\def\spacingset#1{\renewcommand{\baselinestretch}%
{#1}\small\normalsize} \spacingset{1}

%%%%%%%%%%%%%%%%%%%%%%%%%%%%%%%%%%%%%%%%%%%%%%%%%%%%%%%%%%%%%%%%%%%%%%%%%%%%%%

%Selective Borrowing of External Controls in Hybrid Controlled Trials: A Tutorial Using SelectiveIntegrative and intFRT

\if1\anon
{
  \title{\bf Robust Estimation and Inference with Selective Borrowing in Hybrid Controlled Trials: A Tutorial with SelectiveIntegrative and intFRT}

\author[1,2]{Ke Zhu}
\author[1]{Hairong Huang}
\author[1]{Shu Yang}
\author[2]{Xiaofei Wang\footnote{Address for correspondence: Xiaofei Wang, Department of Biostatistics and Bioinformatics, Duke University, Durham, NC 27710, USA. Email: xiaofei.wang@duke.edu}}

\affil[1]{\small Department of Statistics, North Carolina State University, Raleigh, NC 27695, USA}
\affil[2]{Department of Biostatistics and Bioinformatics, Duke University, Durham, NC 27710, USA}
\date{}
\maketitle
} \fi

\if0\anon
{
 \title{\bf Robust Estimation and Inference with Selective Borrowing in Hybrid Controlled Trials: A Tutorial with SelectiveIntegrative and intFRT}
 \date{}
 \author{}
 \maketitle
} \fi

\bigskip
\begin{abstract}
Hybrid controlled trials (HCTs) augment randomized controlled trials (RCTs) with external controls (ECs) to improve statistical efficiency when RCTs face limited sample sizes, slow accrual, or ethical constraints. However, valid use of ECs requires careful adjustment for covariate shift and outcome drift, as inappropriate borrowing may introduce bias and compromise inference. This tutorial provides a practical workflow for estimation and inference in HCTs. We first present a statistical analysis roadmap covering estimands, identification assumptions, eligibility alignment, matching, full and selective borrowing strategies, and both asymptotic inference and randomization tests. We then demonstrate step-by-step implementation using the \texttt{SelectiveIntegrative} and \texttt{intFRT} packages. The workflow is illustrated using a synthetic lung cancer dataset included in the \texttt{intFRT} package that mimics the CALGB 9633 trial and ECs from the National Cancer Database. The tutorial aims to help applied statisticians conduct transparent, interpretable, and reproducible HCT analyses that improve efficiency while maintaining valid inference.
\end{abstract}

\noindent%
\vfill

\newpage
\spacingset{1.8} % DON'T change the spacing!

\section{Introduction}\label{sec-intro}

Randomized controlled trials (RCTs) remain the gold standard for evaluating treatment effects because randomization protects against both measured and unmeasured confounding. However, in many clinical studies, especially rare disease, pediatric, and oncology trials, RCTs may suffer from limited sample sizes, slow accrual, insufficient power, and ethical concerns. Hybrid controlled trials (HCTs), which augment randomized controlled trials (RCTs) with external controls (ECs) from real-world data (RWD) or historical trials, provide a promising strategy for improving statistical efficiency while preserving the RCT population as the target population, given its strong internal validity \citep{mishra2022external,ventz2022design}.

Despite their promise, HCTs introduce statistical challenges. Borrowing ECs can increase precision, but inappropriate borrowing may introduce two types of bias and lead to invalid inference. The first is covariate shift, which arises when observed baseline covariates differ between the RCT and EC populations. The second is outcome drift, also known as hidden bias, which occurs when RCT controls and ECs remain systematically different after adjustment for observed covariates. Over the past decade, emerging Bayesian and frequentist methods have been developed to borrow ECs while mitigating these biases and supporting valid inference. We refer readers to \cite{zhu2026externally} for a review.

This tutorial presents a practical workflow for analyzing HCTs that addresses covariate shift and outcome drift while improving statistical efficiency and maintaining valid inference \citep{gao2025improving,gao2025doubly,zhu2025enhancing,liu2025robust}. To address covariate shift of ECs, we use propensity score matching, weighting, outcome modeling, and doubly robust methods \citep{shan2022simulation,lin2023matching,li2023improving,Valancius2024}. When the full set of ECs may exhibit outcome drift, we use selective borrowing, whose core idea is to identify an outcome-drift-free subset of ECs by comparing each EC individually with the RCT controls in terms of their conditional outcome distributions. In addition to efficient estimation and asymptotic inference, we introduce Fisher randomization tests (FRTs) as a finite-sample exact, model-free, and post-selection-valid inference tool to safeguard type I error rate control. These methods are implemented through two R packages: \texttt{SelectiveIntegrative} and \texttt{intFRT}. We emphasize that selective borrowing should not be viewed as a substitute for careful design-stage assessment of the fitness-for-purpose of external controls. Rather, it serves as an analysis-stage tool that can improve robustness when ECs may be heterogeneous or only partially comparable to the RCT controls.

We first introduce the statistical analysis roadmap in Section~\ref{sec:met}, including estimands, assumptions, RCT-only analysis, eligibility alignment, matching, full borrowing, selective borrowing, and FRTs. Section~\ref{sec:workflow} then provides a step-by-step implementation workflow with reproducible R code and practical implementation details. The workflow is illustrated using a lung cancer application based on the CALGB 9633 trial \citep{Strauss2008} and ECs from the National Cancer Database (NCDB). CALGB 9633 evaluated adjuvant chemotherapy versus observation after surgical resection among patients with Stage IB non-small-cell lung cancer, but the trial was underpowered because of limited sample size and slow accrual. The NCDB provides a large pool of ECs, but differences in covariate distributions and unmeasured variables, such as ECOG performance status, raise concerns about both covariate shift and outcome drift. Because the original patient-level data are not publicly available, this tutorial uses a synthetic dataset included in the \texttt{intFRT} package that mimics the CALGB 9633 and NCDB setting. Section~\ref{sec:con} concludes with future directions.

This tutorial is intended to guide applied statisticians through the key analytic decisions involved in the analysis of HCTs. The proposed workflow should be viewed as a sequence of prespecified steps summarized in the reporting checklist below.

\begin{tcolorbox}[title=Recommended Reporting Checklist for HCT Analyses]
\begin{enumerate}
\item Target estimand and population.
\item EC source and eligibility alignment.
\item Pre-matching strategy and matching ratio.
\item Covariate-overlap diagnostics.
\item RCT-only benchmark results.
\item Full-borrowing results.
\item Selective-borrowing results.
\item FRT results.
\item Sensitivity analyses.
\end{enumerate}
\end{tcolorbox}

\section{A Statistical Analysis Roadmap}
\label{sec:met}

\subsection{Study Objectives, Estimand, and Assumptions}

Our objective is to evaluate the treatment effect of the experimental treatment ($A=1$) versus control ($A=0$). The target population is the RCT population ($S=1$), defined by the trial eligibility criteria. ECs ($S=0$) are incorporated solely to improve statistical efficiency and do not alter the target population.

We consider a binary endpoint, such as 3-year survival status in the motivating example. Let $Y(a)$ denote the potential outcome under treatment $a\in\{0,1\}$ \citep{neyman1923application,rubin1974estimating}. The estimand of interest is the average treatment effect (ATE) in the RCT population,
\[
\tau=\theta_1-\theta_0,
\qquad
\theta_a=\mathbb{E}\{Y(a)\mid S=1\}.
\]

\begin{figure}[t]
    \centering
    \includegraphics[width=0.9\linewidth]{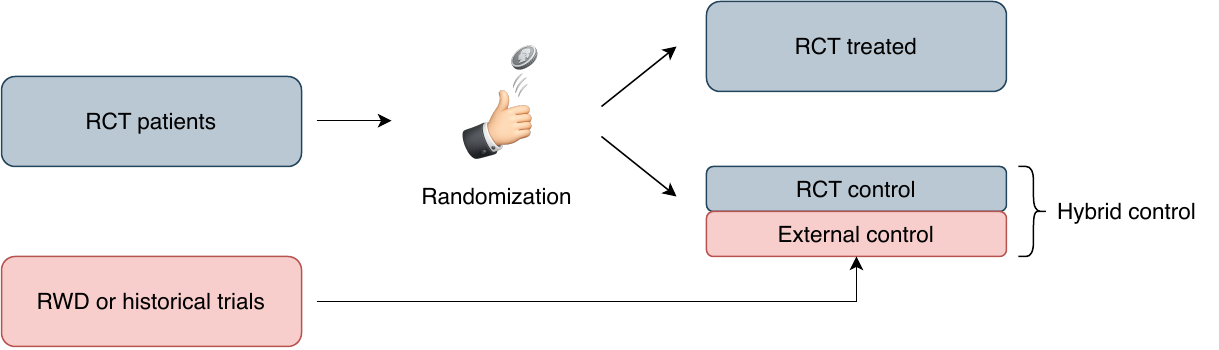}
    \caption{Hybrid controlled trials.}
    \label{fig:hct}
\end{figure}

Within the RCT, treatment is randomized with known allocation probability $\pi_A(x)=\mathbb{P}(A=1\mid S=1,X=x)\equiv\bar\pi_A=n_1/n_{\rm RCT}$, where $n_1$ is the number of RCT-treated patients, $n_0$ is the number of RCT control patients, and $n_{\rm RCT}=n_1+n_0$. In the EC dataset, all $n_{\rm EC}$ patients receive control, so $A=0$ whenever $S=0$.
Figure~\ref{fig:hct} illustrates the structure of an HCT.
Let $\pi_S(x)=\mathbb{P}(S=1\mid X=x)$ denote the sampling propensity score \citep{tipton2013improving}, and let $\bar{\pi}_S=\mathbb{P}(S=1)=n_{\rm RCT}/n$ denote the proportion of RCT patients in the combined dataset, where $n=n_{\rm RCT}+n_{\rm EC}$. Let $Y$ denote the observed outcome. We consider the following identifiability assumptions \citep{li2023improving,Valancius2024}.

\begin{assumption}[RCT identification]
\label{ass:rct}
For units with $S=1$, we assume: (i) the Stable Unit Treatment Value Assumption (SUTVA), $Y=Y(A)$; (ii) treatment unconfoundedness, ${Y(0),Y(1)}\perp A \mid X, S=1$; and (iii) treatment positivity, $0<\pi_A(x)<1$ almost surely.
\end{assumption}

\begin{assumption}[Conditional mean exchangeability]
\label{ass:ec}
$$
\mathbb{E}\{Y(0)\mid S=1,X\}
=
\mathbb{E}\{Y(0)\mid S=0,X\}.
$$
\end{assumption}

\begin{assumption}[Sampling positivity]
\label{ass:pi}
$
\pi_S(x)>0
$
for all $x$ such that $f_X(x)>0$, where $f_X(x)$ denotes the marginal density (or probability mass function) of $X$.
\end{assumption}

Assumption~\ref{ass:rct} is guaranteed by the randomized trial design, and the validity of an RCT-only analysis relies solely on this assumption. Assumption~\ref{ass:ec} requires that EC and RCT control outcomes are comparable after adjustment for observed covariates. We will apply selective borrowing procedures to identify and retain ECs that satisfy Assumption~\ref{ass:ec}. Assumption~\ref{ass:pi} requires that every EC patient has a positive probability of being enrolled in the RCT, although the converse is not required. This assumption is facilitated by restricting ECs to patients who satisfy the RCT eligibility criteria.

%RCT-only Analysis
%build a strong benchmark before borrow EC
%consider good balanced design rather than introduce imbalance
%consider efficient cov adj rather than only difference-in-mean
\textbf{Efficient RCT-only analysis as a benchmark.} Before borrowing ECs, it is important to establish a strong RCT-only benchmark. Any claimed efficiency gain from EC borrowing should be evaluated relative to an efficient RCT-only analysis, so that the gain reflects the true added value of the external data rather than inefficiencies in the RCT analysis itself. This includes using a well-balanced randomization design \citep{rosenberger2008handling} and efficient covariate adjustment methods \citep{FDA2023CovariateAdjustment} rather than relying solely on unadjusted difference-in-means estimators.

\textbf{Two sources of bias.} When borrowing ECs, two sources of bias should be considered. The first is \textbf{covariate shift}, arising from differences in the distribution of observed covariates between the RCT and EC populations, which can be addressed through matching, weighting, outcome regression, or doubly robust methods. The second is \textbf{outcome drift}, which occurs when RCT controls and ECs remain \emph{systematically different} after adjustment for observed covariates, violating Assumption~\ref{ass:ec}. It may arise from unmeasured confounders, differences in clinical practice, or temporal changes \citep{FDA2023}. For example, ECOG performance status is available in CALGB 9633 but not in NCDB.

\subsection{Eligibility Alignment and Matching}
%matching itself is not enough
%but better than no matching; usually improve stability

For eligibility alignment, analysts should verify that ECs satisfy the RCT inclusion/exclusion criteria, harmonize covariate and endpoint definitions, restrict ECs to the RCT covariate support, and assess overlap using graphical diagnostics and numerical measures such as standardized mean differences and sampling-score distributions \citep{austin2009balance,greifer2020covariate}. ECs with non-harmonizable endpoints, follow-up rules, or covariate values outside the RCT support should generally be excluded before borrowing.

According to Assumption~\ref{ass:pi}, we restrict the EC cohort to patients whose covariate values fall within the support of the RCT population. This is sufficient because the target population is the RCT population. Therefore, we do not require the support of the RCT covariates to coincide with that of the EC cohort. For example, if patients with tumor sizes smaller than 3 cm are present in the RCT but absent from the EC cohort, treatment effects for these patients can still be estimated using the randomized treated and control groups within the RCT. In contrast, EC patients with covariate values outside the RCT support cannot contribute to inference for the target population and are therefore excluded.

Beyond covariate range restrictions, we perform nearest-neighbor matching based on the sampling propensity score to select a subset of ECs that more closely resembles the covariate distribution of the RCT population \citep{ho2011matchit}. We consider $K{:}1$ matching, with $K=1$ by default, although alternative values can be examined in sensitivity analyses.

Matching alone is generally insufficient to eliminate bias due to covariate shift because exact covariate balance is rarely achieved. Nevertheless, pre-matching is often beneficial for two reasons. First, it can improve the stability of downstream analyses \citep{shan2022simulation,Qiu2025}, and reduce sensitivity to model selection and model misspecification \citep{guo2023statistical}. Second, EC datasets are often substantially larger than the RCT. For example, in our motivating study, the EC cohort contains 11,700 patients whereas the RCT includes only 335 patients. Pre-matching restricts the EC sample to a prespecified size relative to the RCT, preventing the EC data from dominating the analysis.

\subsection{Borrowing, Estimation, and Asymptotic Inference}
\subsubsection{Full Borrowing}
\label{sec:fb}
Let $n_{\rm EC}$ denote the number of ECs retained after eligibility alignment and pre-matching. We first consider \textbf{full borrowing} (FB), where the full set of matched ECs is included in the analysis. The term ``full'' refers to borrowing from the full set of matched ECs retained after preprocessing, rather than all ECs in the original dataset, and contrasts with selective borrowing, which uses only a subset of matched ECs.
Full borrowing addresses residual covariate shift through confounding adjustment methods such as inverse probability weighting (IPW), stabilized inverse probability weighting (sIPW), calibration weighting (CW), outcome modeling (OM), augmented inverse probability weighting (AIPW), and augmented calibration weighting (ACW), while relying on Assumption~\ref{ass:ec} to rule out outcome drift. Among these methods, we focus on AIPW as the primary estimator in this section. We will demonstrate the implementation of all methods in Section~\ref{sec:workflow} and refer readers to \cite{Valancius2024} and \cite{liu2025robust} for mathematical details.

AIPW combines outcome modeling and sampling propensity score modeling and enjoys the \emph{double robustness} property, meaning that the estimator remains consistent if either the outcome model or the sampling propensity score model is correctly specified. Because ECs are available only for the control arm, we estimate $\theta_1$ and $\theta_0$ separately.

For the treatment arm mean $\theta_1$, estimation relies solely on the RCT:
\[
\widehat{\theta}_1
=
\frac{1}{n_{\rm RCT}}
\sum_{i=1}^{n}
S_i
\left[
\widehat{\mu}_{1}(X_i)
+
\frac{A_i}{\pi_A(X_i)}
\left\{
Y_i-\widehat{\mu}_{1}(X_i)
\right\}
\right],
\]
where
$
\mu_1(x)
=
\mathbb{E}(Y\mid A=1,S=1,X=x)
$
denotes the conditional mean outcome among treated RCT patients and $\widehat{\mu}_{1}(x)$ is its estimator. The first term predicts outcomes under treatment using the outcome model, while the second term corrects for potential model misspecification using the known randomization probability $\pi_A(x)$.

For the control arm mean $\theta_0$, we borrow information from both RCT controls and ECs. Let
$
\mu_0(x)
=
\mathbb{E}(Y\mid A=0,X=x)
$
denote the conditional mean outcome under control and let $\widehat{\mu}_0(x)$ denote its estimator. Under Assumption~\ref{ass:ec}, RCT controls and ECs share the same conditional mean outcome given $X$, allowing both sources to be pooled for estimation:
\begin{equation}
\label{eq:aipw}
\widehat{\theta}_{0}^{\rm FB\mbox{-}AIPW}
=
\frac{1}{n_{\rm RCT}}
\sum_{i=1}^{n}
\left[
S_i\widehat{\mu}_0(X_i)
+
\frac{
\widehat{\pi}_S(X_i)
\left\{
(1-A_i)S_i+(1-S_i)\widehat r(X_i)
\right\}
}
{
\{1-\pi_A(X_i)\}\widehat{\pi}_S(X_i)
+
\{1-\widehat{\pi}_S(X_i)\}\widehat r(X_i)
}
\left\{
Y_i-\widehat{\mu}_0(X_i)
\right\}
\right],
\end{equation}
where $\widehat{\pi}_S(x)$ is the estimated sampling propensity score and
\[
r(x)
=
\frac{
\mathrm{Var}(Y\mid S=1,A=0,X=x)
}{
\mathrm{Var}(Y\mid S=0,A=0,X=x)
}
\]
is the conditional variance ratio between RCT controls and ECs, with $\widehat r(x)$ denoting its estimator. The weighting term adjusts for residual covariate shift between the RCT and EC populations, while the variance ratio allows efficient combination of information from the two control sources \citep{li2023improving}.

Finally, the ATE in the RCT population is estimated by
\[
\widehat{\tau}^{\rm FB\mbox{-}AIPW}
=
\widehat{\theta}_1
-
\widehat{\theta}_0^{\rm FB\mbox{-}AIPW}.
\]
Statistical inference can be conducted using sandwich variance estimators or bootstrap variance estimators together with asymptotic normality.

\subsubsection{Penalized Selective Borrowing}

To further address partial violations of Assumption~\ref{ass:ec}, we consider \textbf{selective borrowing}, which uses the RCT controls as a benchmark to identify a subset of ECs that are compatible with Assumption~\ref{ass:ec}. We consider two selective borrowing approaches.

\textbf{Penalized selective borrowing} (PSB) estimates the individual bias of each EC relative to the RCT controls and uses penalization methods, such as the adaptive lasso \citep{zou2006adaptive}, to shrink small estimated biases toward zero, thereby encouraging borrowing from compatible ECs \citep{gao2025improving}.
Specifically, for each EC, PSB defines an individual bias parameter
\[
b_i
=
\mathbb{E}(Y\mid S=0,A=0, X_i)
-
\mathbb{E}(Y\mid S=1,A=0, X_i),
\qquad i:S_i=0,
\]
which measures the degree of outcome drift relative to the RCT controls. A nonzero value of $b_i$ indicates outcome incompatibility between the EC and RCT control populations.
The procedure consists of three steps:

\begin{enumerate}

\item \textbf{Initial bias estimation.}
Fit separate outcome models in the EC and RCT control groups and estimate
$\widehat b_i
=
\widehat\mu_{0,EC}(X_i)
-
\widehat\mu_{0,RCT}(X_i),
$
where
$
\mu_{0,EC}(x)
=
\mathbb E(Y\mid S=0,A=0,X=x)$,
$
\mu_{0,RCT}(x)
=
\mathbb E(Y\mid S=1,A=0,X=x).
$
Let
$\widehat{\boldsymbol b}
=
(\widehat b_1,\ldots,\widehat b_{n_{\rm EC}})^\T$
denote the vector of estimated biases.

\item \textbf{Penalization.}
Refine the estimated biases by solving
\[
\widetilde{\boldsymbol b}
=
\arg\min_{\boldsymbol b}
\left\{
(\widehat{\boldsymbol b}-\boldsymbol b)^\T
\widehat{\Sigma}_b^{-1}
(\widehat{\boldsymbol b}-\boldsymbol b)
+
\lambda
\sum_{i=1}^{n_{\rm EC}}
\frac{|b_i|}
{|\widehat b_i|^\nu}
\right\},
\]
where $\widehat{\Sigma}_b$ is the estimated covariance matrix of $\widehat{\boldsymbol b}$, and $\lambda$ and $\nu$ are tuning parameters.

\item \textbf{Construct the selected EC set.}
Define the selected EC subset as
\[
\widehat{\mathcal E}_{\rm PSB}
=
\{i:\widetilde b_i=0\},
\]
and apply the full-borrowing estimator in Section~\ref{sec:fb} with the selected ECs.

\end{enumerate}

Under selection consistency, meaning that the selected ECs asymptotically satisfy Assumption~\ref{ass:ec}, the AIPW (or ACW) estimator constructed using the selected ECs yields a selective-borrowing estimator together with valid asymptotic inference.

\subsubsection{Conformal Selective Borrowing}

\textbf{Conformal selective borrowing} (CSB) uses conformal $p$-values \citep{vovk2005algorithmic,angelopoulos2024theoretical} to assess the \emph{individual exchangeability} of each EC patient with the RCT control population and selects ECs using a data-adaptive threshold chosen to minimize the estimated mean squared error (MSE) of the resulting estimator \citep{zhu2025enhancing,liu2025robust}.
Individual exchangeability means that the joint distribution of the RCT control patients and a given EC patient remains unchanged under permutations of their ordering \citep{angelopoulos2024theoretical}. This assumption is generally stronger than Assumption~\ref{ass:ec}, which requires only conditional mean exchangeability, i.e., equality of conditional outcome means between the EC and RCT control populations given covariates.
To compute the conformal $p$-value for an EC subject $i$, we proceed as follows:

\begin{enumerate}

\item \textbf{Sample splitting.}
Randomly split the RCT controls into a training set $\mathcal T$ and a calibration set $\mathcal C$. In practice, cross-validation is recommended to improve data utilization \citep{barber2021predictive,zhu2025enhancing,liu2025robust}.

\item \textbf{Compute conformal scores.}
Measure the compatibility of the EC with the RCT controls using a conformal score. For a continuous outcome, we can use the absolute residual (AR) score
\[
s_i=|Y_i-\widehat f(X_i)|,
\]
where $\widehat f(x)$ is an outcome prediction model fitted using RCT controls in $\mathcal T$. In practice, conformalized quantile regression is recommended as the conformal score \citep{romano2019conformalized,zhu2025enhancing}.

For a binary outcome, we can use the nearest-neighbor conformal score
\[
s_i
=
\min\{d(X_i,X_l):Y_l=Y_i,\; l\in\mathcal T\},
\]
where $d(\cdot,\cdot)$ is a distance metric (e.g., Euclidean distance). Larger conformal scores indicate lower compatibility with the RCT controls.

\item \textbf{Calibration.}
For each calibration subject $k\in\mathcal C$, compute the corresponding conformal score $s_k$. The conformal $p$-value for EC subject $i$ is
\[
p_i
=
\frac{
\sum_{k\in\mathcal C}
\mathbb{I}(s_k\ge s_i)+1
}
{|\mathcal C|+1}.
\]
In practice, label-conditional conformal $p$-values are generally recommended for binary outcomes \citep{liu2025robust}.
Small conformal $p$-values indicate evidence of outcome drift, whereas large conformal $p$-values suggest compatibility with the RCT controls.

\end{enumerate}

With conformal $p$-values, the CSB procedure consists of three steps:

\begin{enumerate}

\item \textbf{Compute conformal $p$-values.}
Compute conformal $p$-values $\{p_i:S_i=0\}$ for all ECs using the procedure described above.

\item \textbf{Construct a class of selective-borrowing estimators.}
Given a threshold $\gamma\in[0,1]$, define the selected EC subset
\[
\widehat{\mathcal E}_{\rm CSB}(\gamma)
=
\{i:p_i>\gamma\}.
\]
Apply the full-borrowing estimator described in Section~\ref{sec:fb} using only ECs in $\widehat{\mathcal E}_{\rm CSB}(\gamma)$ to obtain a selective-borrowing estimator $\widehat\tau_\gamma$. Two important special cases are $\gamma=0$, which corresponds to the full-borrowing estimator using all ECs, and $\gamma=1$, which corresponds to the no-borrowing estimator using only the RCT data.

\item \textbf{Select the optimal threshold.}
Choose the threshold that minimizes the estimated mean squared error (MSE),
\[
\widehat\gamma
=
\arg\min_{\gamma}
\widehat{\rm MSE}(\widehat\tau_\gamma).
\]
Following \cite{zhu2025enhancing}, the no-borrow estimator is treated as approximately unbiased for the target treatment effect and serves as a benchmark for estimating the bias component of $\widehat{\rm MSE}(\widehat\tau_\gamma)$, while variance components are estimated using sandwich variance estimators or bootstrap resampling for a fixed $\gamma$. The final CSB estimator is then constructed using the selected EC set
$
\widehat{\mathcal E}_{\rm CSB}(\widehat\gamma)
=
\{i:p_i>\widehat\gamma\}.
$

\end{enumerate}

\subsection{Fisher Randomization Test}

Although asymptotic inference provides a useful statistical evaluation of treatment effects in HCTs, it relies on following conditions. First, it relies on large-sample normal approximations, which may be inaccurate when the RCT sample size is small. Second, doubly robust methods require either the outcome model or the sampling propensity score model to be correctly specified and may perform poorly when both models are misspecified. Third, FB relies on Assumption~\ref{ass:ec}, while PSB and CSB rely on selection consistency.

Fisher randomization tests (FRTs) provide a complementary inference framework with finite-sample exact type I error control in HCTs \citep{fisher1935}. Unlike asymptotic inference, which typically targets the \emph{weak null hypothesis} of zero average treatment effect, FRTs test the \emph{sharp null hypothesis} that the individual treatment effect is zero for every RCT participant. 
FRTs condition on the potential outcomes and covariates, treating the treatment assignment mechanism as the sole source of randomness and the EC data as fixed.
FRTs are model-free, do not rely on large-sample approximations, and remain valid after adaptive EC selection when the selection procedure is repeated within each randomization permutation. Rejection of the sharp null indicates evidence of a treatment effect, for at least some patients, even when the average treatment effect is zero. Therefore, FRTs are particularly attractive in exploratory Phase II studies, rare disease settings, and biomarker-driven trials with potentially heterogeneous treatment effects. We refer readers to \cite{berger2000pros,simon2011using,carter2024regulatory} for further discussion of FRTs and permutation tests in clinical trials and regulatory settings.

Under the sharp null hypothesis
\[
H_0^{\rm sharp}: Y_i(1)=Y_i(0), \qquad i=1,\ldots,n_{\rm RCT},
\]
all missing potential outcomes for RCT participants can be imputed from the observed outcomes. Let $T(\cdot)$ denote a test statistic of interest, such as the CSB-AIPW estimator. The FRT proceeds as follows:

\begin{enumerate}
\item Compute the observed test statistic $T(\mathbf{A}^{\rm obs})$ using the actual treatment assignments.

\item For $b=1,\ldots,B$, generate a permuted treatment assignment $\mathbf{A}^{(b)}$ for RCT participants according to the original randomization procedure. The treatment assignments of ECs remain fixed because they were not randomized.

\item Reconstruct the entire HCT analysis under $\mathbf{A}^{(b)}$, including EC selection (if applicable) and estimation, and compute the corresponding test statistic $T(\mathbf{A}^{(b)})$.

\item Estimate the $p$-value as
\[
\hat p^{\rm FRT}
=
\frac{
\sum_{b=1}^{B}
\mathbb{I}\!\left\{
|T(\mathbf{A}^{(b)})|
\ge
|T(\mathbf{A}^{\rm obs})|
\right\}+1
}
{B+1}.
\]
\end{enumerate}

With $B$ permutations, the Monte Carlo standard error of the estimated p-value is approximately $\sqrt{p^{\rm FRT}(1-p^{\rm FRT})/B}$, where $p^{\rm FRT}=\mathbb{P}_{\mathbf{A}^{(b)}}(|T(\mathbf{A}^{(b)})|
\ge
|T(\mathbf{A}^{\rm obs})|)$; therefore, a larger $B$ may be required when the p-value is close to the significance threshold.

By replaying the entire analysis under treatment assignments generated from the original randomization mechanism, FRT directly approximates the randomization distribution of the test statistic. As a result, it maintains valid type I error control even when FB-AIPW is applied under violations of Assumption~\ref{ass:ec}, or when CSB-AIPW involves additional uncertainty from the EC selection procedure. Nevertheless, when substantial outcome drift is present, the power of FRT using FB-AIPW as the test statistic can be substantially lower than that of an RCT-only analysis. Therefore, we recommend using CSB-AIPW as the test statistic within FRT, as selective borrowing can improve power by preferentially incorporating ECs that are compatible with the RCT controls.

\vspace{-15pt}

\section{Workflow Illustration}
\label{sec:workflow}

\subsection{Import Data}

We illustrate the workflow using a synthetic data set included in the \texttt{intFRT} package. The data were generated to resemble the CALGB 9633 trial and the NCDB external control cohort, allowing users to reproduce the tutorial directly. We also include a complete R Markdown file as Supplementary Material, with technical details on package installation, runtime recording, and code for generating the plots.

\begin{rcode}
library(tidyverse)
library(MatchIt)
library(intFRT)
library(SelectiveIntegrative)
data("lungcancer", package = "intFRT")
data("lungcancer_truth", package = "intFRT")
res_time <- 3
lungcancer2 <- lungcancer %>%
  left_join(lungcancer_truth, by = "patid") %>%
  mutate(
    survtime_uncen = ifelse(treat == 1, T1, T0),
    y = case_when(
      survtime_uncen > res_time ~ 1L,
      survtime_uncen <= res_time ~ 0L
    )
  )
\end{rcode}

Because this is a synthetic dataset, the package also includes the potential outcomes (\texttt{lungcancer\_truth}) used to generate the observed outcomes. In practice, analysts would start from the observed RCT and EC datasets below (\texttt{data\_rct0} and \texttt{data\_ec0}).

\begin{rcode}
dat_rct0 <- lungcancer2 %>%
  filter(cohort == "C9633") %>%
  select(treat, sex, age, race, hist, tsize, y) %>%
  mutate(sample = 1)
dat_ec0 <- lungcancer2 %>%
  filter(cohort == "EHR", treat == 0) %>%
  select(treat, sex, age, race, hist, tsize, y) %>%
  mutate(sample = 0)
\end{rcode}

We consider a binary outcome indicating 3-year survival (\texttt{y = 1} for survival and \texttt{y = 0} for death). The treatment indicator is defined as \texttt{treat = 1} for adjuvant chemotherapy after surgery and \texttt{treat = 0} for observation after surgery. The data include an RCT cohort (\texttt{sample = 1}), an external control cohort (\texttt{sample = 0}), and five baseline covariates.

\begin{rcode}
dat_rct0 %>% slice_sample(n = 3)
\end{rcode}

\begin{verbatim}
##    treat sex age race hist tsize y sample
## 1      0   1  63    1    0   2.5 1      1
## 2      0   1  74    1    0   7.0 0      1
## 3      0   1  57    1    0   8.1 0      1
\end{verbatim}

\begin{rcode}
dat_ec0 %>% slice_sample(n = 3)
\end{rcode}

\begin{verbatim}
##    treat sex age race hist tsize y sample
## 1      0   0  84    1    0   6.3 1      0
## 2      0   0  73    0    0   3.5 1      0
## 3      0   1  78    1    1   5.3 0      0
\end{verbatim}

\subsection{Eligibility Alignment and Matching}

We restrict ECs to the support of the two continuous covariates in the RCT, as the remaining three binary covariates are represented in both datasets. More generally, analysts should also assess categorical covariate support, missing-data patterns, calendar-time compatibility, treatment and endpoint definitions, and data quality differences across sources.

\begin{figure}[t]
    \centering
    \includegraphics[width=1\linewidth]{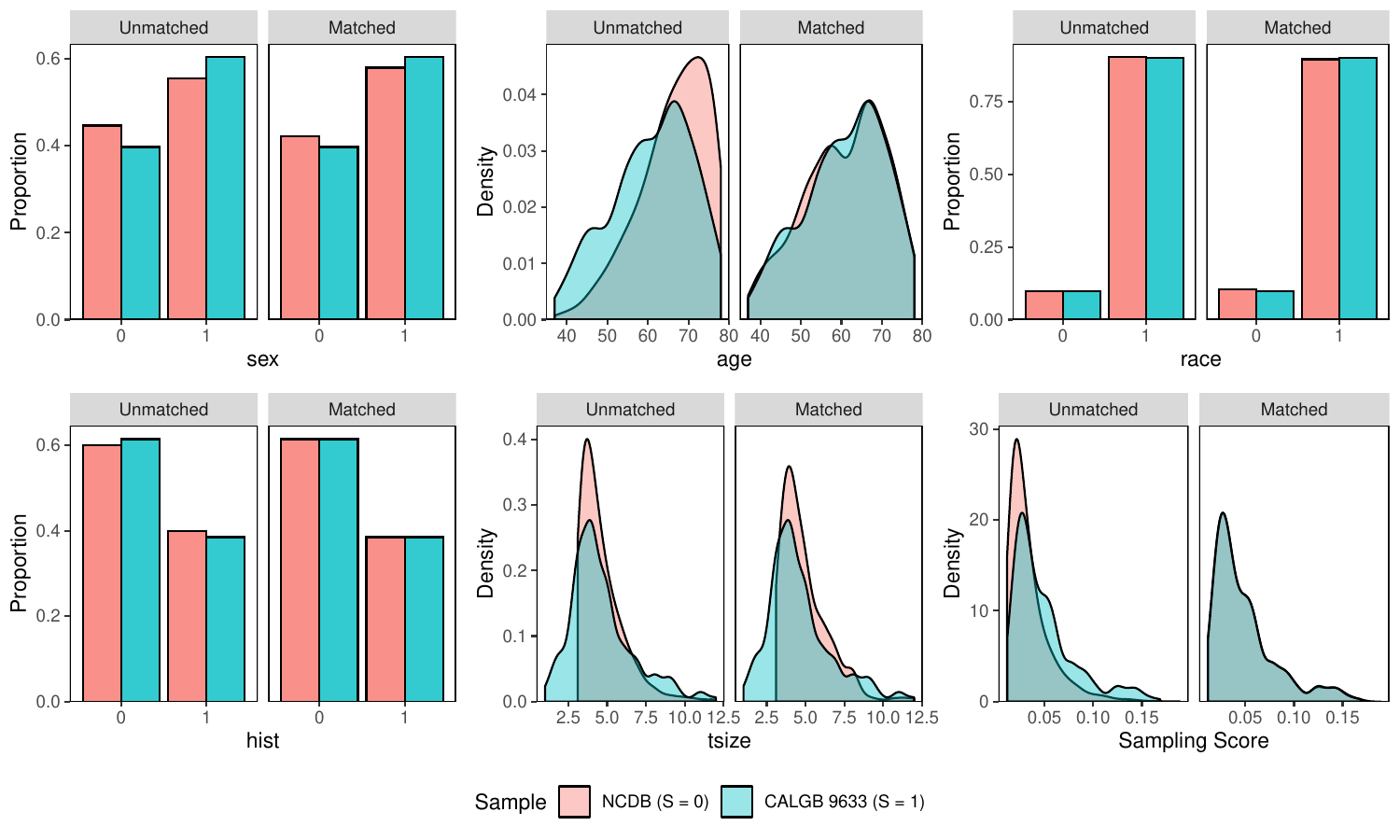}
    \caption{Covariate balancing before and after matching.}
    \label{fig:match}
\end{figure}

We then perform propensity score matching using all five covariates. The resulting matched data set is used in all subsequent analyses. Figure~\ref{fig:match} displays covariate balance before and after matching. The figure shows that (i) the ECs lie within the covariate support of the RCT, (ii) covariate balance is substantially improved after matching, and (iii) some residual imbalance remains, motivating further adjustment in the analysis stage.

\begin{rcode}
# 1. Restrict ECs to the RCT support
dat_restrict <- bind_rows(
  dat_rct0,
  dat_ec0 %>%
    filter(
      between(age, min(dat_rct0$age), max(dat_rct0$age)),
      between(tsize, min(dat_rct0$tsize), max(dat_rct0$tsize))
    )
)
# 2. Perform matching
m.out <- matchit(
  sample ~ sex + age + race + hist + tsize,
  data = dat_restrict,
  method = "nearest",
  distance = "glm",
  replace = FALSE,
  exact = "hist"
)
dat <- match.data(m.out) %>% select(-weights, -subclass, -distance)
\end{rcode}

\subsection{Borrowing, Estimation, and Asymptotic Inference}

After eligibility alignment and matching, the resulting matched dataset is ready for analysis; see Table~\ref{tab:char} for a summary of patient characteristics. The code below uses a standard interface to define the data inputs and working models required for the subsequent borrowing and inference procedures.

\begin{table}[p]

\caption{Patient characteristics in the synthetic dataset mimicking CALGB 9633 and NCDB after eligibility alignment and matching.}
\centering
\label{tab:char}
\begin{tabular}[t]{lcccc}
\toprule
\textbf{Characteristic} & \makecell[c]{\textbf{RCT treated$^*$}\ \ \\N = 167} & \makecell[c]{\textbf{RCT controls}\ \ \\N = 168} & \makecell[c]{\textbf{Matched ECs}\ \ \\N = 335} & \makecell[c]{\textbf{Overall}\ \ \\N = 670}\\
\midrule
Sex &  &  &  & \\
\hspace{1em}0 & 69 (41\%) & 64 (38\%) & 141 (42\%) & 274 (41\%)\\
\hspace{1em}1 & 98 (59\%) & 104 (62\%) & 194 (58\%) & 396 (59\%)\\
Age & 60.53 (10.36) & 61.46 (9.76) & 60.94 (10.06) & 60.97 (10.05)\\
Race &  &  &  & \\
\addlinespace
\hspace{1em}0 & 18 (11\%) & 15 (8.9\%) & 35 (10\%) & 68 (10\%)\\
\hspace{1em}1 & 149 (89\%) & 153 (91\%) & 300 (90\%) & 602 (90\%)\\
Histology &  &  &  & \\
\hspace{1em}0 & 99 (59\%) & 107 (64\%) & 206 (61\%) & 412 (61\%)\\
\hspace{1em}1 & 68 (41\%) & 61 (36\%) & 129 (39\%) & 258 (39\%)\\
\addlinespace
Tumor size & 4.69 (2.23) & 4.59 (1.86) & 4.90 (1.51) & 4.77 (1.80)\\
3-year success & 133 (80\%) & 124 (74\%) & 227 (68\%) & 484 (72\%)\\
\bottomrule
\multicolumn{5}{l}{\rule{0pt}{1em}\textsuperscript{*} n (\%); Mean (SD)}\\
\end{tabular}
\end{table}

\begin{rcode}
# data separation
dat_rct <- dat %>% filter(sample == 1)
dat_ec <- dat %>% filter(sample == 0)
# set data object with model specification
dat_obj <- intFRT::dataInput(
  dat_rct, dat_ec,
  models = list(
    outcome = y ~ (sex + age + race + hist + tsize) * treat, # outcome model
    ps = treat ~ 1, # treatment propensity score
    ss = ~ sex + age + race + hist + tsize, # sampling propensity score
    cf = ~ sex + age + race + hist + tsize # conformal score model
  )
)
\end{rcode}

\subsubsection{No Borrowing (RCT-only analysis)}

As a benchmark, we conduct an RCT-only analysis without borrowing ECs. The function \texttt{ec\_borrow()} takes the pre-defined \texttt{dat\_obj} as input, with \texttt{outcome.type = "bin"} indicating a binary outcome. We use the covariate-adjusted estimator (\texttt{method = "NbCovAdj"}) to improve efficiency relative to an unadjusted analysis.
The output includes the treatment effect estimate (\texttt{tau.hat}), its estimated standard error (\texttt{sd.hat}), the asymptotic confidence interval limits (\texttt{ci.lower} and \texttt{ci.upper}), the asymptotic $p$-value (\texttt{p.value}), the number of borrowed ECs (\texttt{n.sel}), and the effective sample size of the borrowed ECs (\texttt{ess.sel}).

\begin{rcode}
result_nb <- ec_borrow(data.rct = dat_obj[[1L]], data.ec = dat_obj[[2L]],
  outcome.type = "bin", method = "NbCovAdj")
\end{rcode}

% \begin{verbatim}
% ## # A tibble: 3 x 8
% ##   tau   tau.hat sd.hat ci.lower ci.upper p.value n.sel ess.sel
% ##   <chr>   <dbl>  <dbl>    <dbl>    <dbl>   <dbl> <int>   <int>
% ## 1 RD     0.0568 0.0474  -0.0361    0.150   0.231     0       0
% ## 2 RR     1.08   0.0668   0.953     1.22    0.233     0       0
% ## 3 OR     1.38   0.369    0.814     2.33    0.232     0       0
% \end{verbatim}

\subsubsection{Full Borrowing}

We next consider Full Borrowing (FB), which incorporates all matched ECs into the analysis. The package implements six estimators that address covariate shift using different adjustment strategies: OM, IPW, sIPW, CW, ACW, and AIPW. 
We illustrate all six methods below and use FB-AIPW as the representative approach in subsequent comparisons. Results for the remaining five methods are provided in the Supplementary Materials.
For OM, IPW, sIPW, and CW, variance estimation is based on the bootstrap. We therefore set \texttt{n.boot = 1000} for 1,000 bootstrap replicates and enable parallel computing with \texttt{n.cores = 10L}.
Since \texttt{outcome.type = “bin”}, the function reports results for three treatment effect estimands: risk difference (RD), risk ratio (RR), and odds ratio (OR). For simplicity, we focus on RD in the comparisons below.

\begin{rcode}
ec_borrow(data.rct = dat_obj[[1L]], data.ec = dat_obj[[2L]],
          outcome.type = "bin", method = "OM", n.boot = 1000L, n.cores = 10L)
ec_borrow(data.rct = dat_obj[[1L]], data.ec = dat_obj[[2L]],
          outcome.type = "bin", method = "IPW", n.boot = 1000L, n.cores = 10L)
ec_borrow(data.rct = dat_obj[[1L]], data.ec = dat_obj[[2L]],
          outcome.type = "bin", method = "sIPW", n.boot = 1000L, n.cores = 10L)
ec_borrow(data.rct = dat_obj[[1L]], data.ec = dat_obj[[2L]],
          outcome.type = "bin", method = "CW", n.boot = 1000L, n.cores = 10L)
ec_borrow(data.rct = dat_obj[[1L]], data.ec = dat_obj[[2L]],
          outcome.type = "bin", method = "ACW")
# use FB-AIPW as the representative approach
result_fb <- ec_borrow(
  data.rct = dat_obj[[1L]], data.ec = dat_obj[[2L]],
  outcome.type = "bin", method = "AIPW"
)
\end{rcode}

\subsubsection{Conformal Selective Borrowing}
\label{sec:csb-code}

We implement CSB in this subsection by specifying \texttt{method = "CfAIPW"}. We set \texttt{outcome.type = "bin"} and recommend using a nearest-neighbor conformal score with label-conditional conformal p-values through the \texttt{cf.regressor} and \texttt{cf.control} arguments shown below. The key tuning parameter is $\gamma$, which controls the amount of borrowing, with smaller values leading to more borrowing. We select $\gamma$ by minimizing an empirical MSE criterion that balances bias and variance relative to the NB estimator. For a prospective analysis, the candidate grid for $\gamma$, the conformal score, the sample-splitting or cross-fitting scheme, and the MSE criterion should be prespecified.

\begin{rcode}
# No Borrow benchmark
tauhat_nb <- result_nb$NbCovAdj$tau.hat["RD"] %>% unname()
# empirical MSE for various gamma
gamma_seq <- seq(0, 1, by = 0.1)
gamma_seq[11] <- 1 + 1e-8
MSE_seq <- map_dbl(gamma_seq, function(gamma) {
  result_csb <- ec_borrow(
    dat_obj[[1L]],
    dat_obj[[2L]],
    outcome.type = "bin",
    method = "CfAIPW",
    gamma.select = gamma,
    seed = 1234,
    cf.regressor = Regressor("NN", list(k = 1)),
    cf.control = list(score = "NN", label.conditional = TRUE)
  )
  tauhat_gamma <- result_csb$CfAIPW$tau.hat["RD"] %>% unname()
  sd_gamma <- result_csb$CfAIPW$sd.hat["RD"] %>% unname()
  bias2 <- (tauhat_gamma - tauhat_nb)^2
  bias2 + sd_gamma^2
})
# gamma plot
plot(gamma_seq, MSE_seq, type = "b")
# gamma with minimal MSE
gamma_hat <- gamma_seq[which.min(MSE_seq)]
\end{rcode}

% \begin{verbatim}
% ## [1] 0.6
% \end{verbatim}

\begin{figure}[p]
    \centering
    \includegraphics[width=0.9\linewidth]{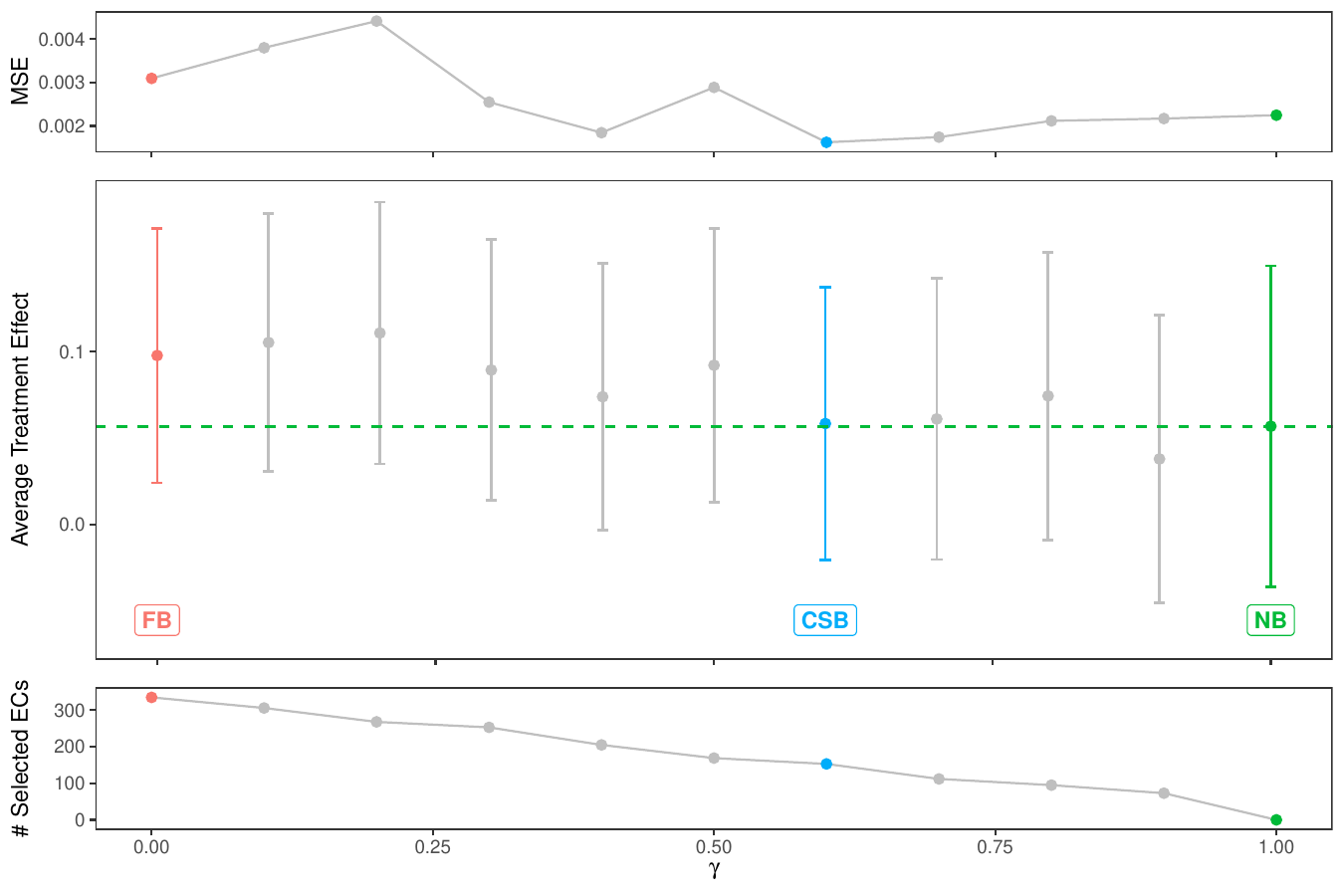}
    \caption{Sensitivity analysis across different values of $\gamma$. NB = no borrowing; FB = full borrowing; CSB = conformal selective borrowing.}
    \label{fig:gamma}
\end{figure}

The empirical MSE is minimized at $\hat{\gamma}=0.6$, as shown in the top panel of Figure~\ref{fig:gamma}. We then run CSB using this value while keeping all other arguments unchanged.
Note that the following code performs only estimation and asymptotic inference using CSB by leaving the argument \texttt{n.fisher} at its default value of \texttt{NULL}. As a result, FRT is not performed, avoiding the additional computational burden associated with permutation.

\begin{rcode}
result_csb <- ec_borrow(
  dat_obj[[1L]], dat_obj[[2L]],
  outcome.type = "bin",
  method = "CfAIPW",
  gamma.select = gamma_hat,
  seed = 1234,
  cf.regressor = Regressor("NN", list(k = 1)),
  cf.control = list(score = "NN", label.conditional = TRUE)
)
\end{rcode}

% We compare NB-AIPW, FB-AIPW, and CSB-AIPW using the RD as the estimand.

% \begin{rcode}
% result_rd <- bind_rows(
%   result_nb,
%   result_fb,
%   result_csb
% ) %>%
%   filter(tau == "RD") %>%
%   select(-tau) %>%
%   mutate(
%     method = c("NB-AIPW", "FB-AIPW", "CSB-AIPW"),
%     .before = everything()
%   )
% \end{rcode}

% \begin{verbatim}
% ## # A tibble: 3 x 8
% ##   method   tau.hat sd.hat ci.lower ci.upper p.value n.sel ess.sel
% ##   <chr>      <dbl>  <dbl>    <dbl>    <dbl>   <dbl> <int>   <dbl>
% ## 1 NB AIPW   0.0568 0.0474  -0.0361    0.150 0.231       0      0 
% ## 2 FB AIPW   0.0978 0.0376   0.0241    0.172 0.00932   335    331.
% ## 3 CSB AIPW  0.0584 0.0402  -0.0205    0.137 0.147     153    140.
% \end{verbatim}

\begin{table}[t]
\centering
\caption{Estimation and inference results from three borrowing strategies.}
\label{tab:hct_result}
\begin{threeparttable}
\begin{tabular}{lccccccc}
\toprule
Method & Estimate & SE & 95\% CI &
Asym. $p$ & FRT $p$ & $n_{\rm borrow}$ & ESS \\
\midrule
NB-AIPW  & 0.0568 & 0.0474 & $(-0.0361,\;0.1497)$ & 0.2310 & 0.227 &   0 &   0 \\
FB-AIPW  & 0.0978 & 0.0376 & $( 0.0241,\;0.1715)$ & 0.0093 & 0.120 & 335 & 331 \\
CSB-AIPW & 0.0584 & 0.0402 & $(-0.0205,\;0.1373)$ & 0.1470 & 0.187 & 153 & 140 \\
\bottomrule
\end{tabular}

\begin{tablenotes}[flushleft]
\footnotesize
\item NB = no borrowing; FB = full borrowing; CSB = conformal selective borrowing; AIPW = augmented inverse probability weighting.
\item Estimate = treatment effect estimate (\texttt{tau.hat}); SE = estimated standard error (\texttt{sd.hat}); 95\% CI = asymptotic 95\% confidence interval (\texttt{ci.lower} and \texttt{ci.upper}); Asym.\ $p$ = asymptotic $p$-value (\texttt{p.value}); FRT $p$ = Fisher randomization test $p$-value (\texttt{FRT\_p.value}); $n_{\rm borrow}$ = number of borrowed external controls (\texttt{n.sel}); ESS = effective sample size of the borrowed external controls (\texttt{ess.sel}).
\end{tablenotes}

\end{threeparttable}
\end{table}

\begin{figure}[p]
    \centering
    \includegraphics[width=0.9\linewidth]{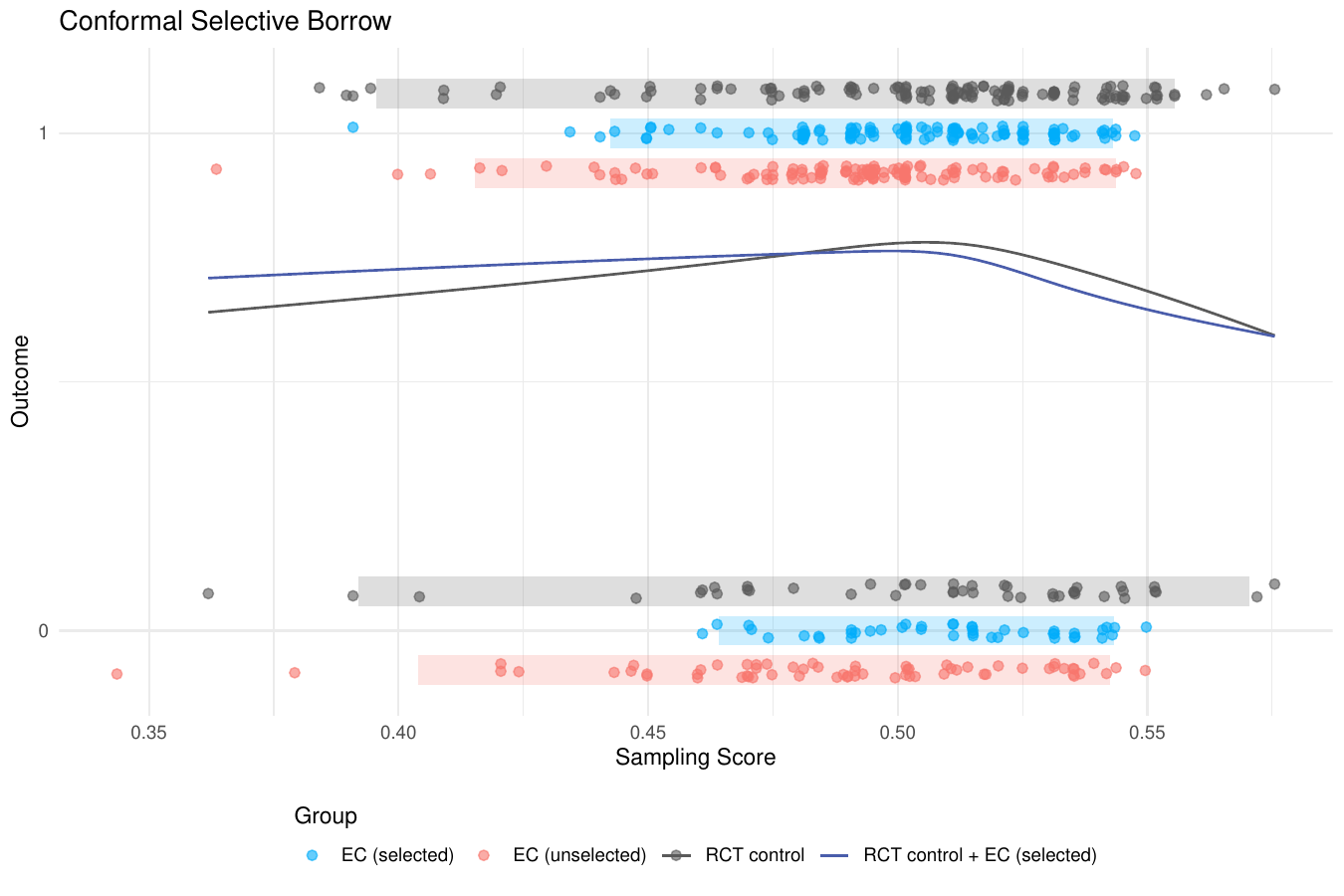}
    \caption{Selected ECs via CSB.}
    \label{fig:sel}
\end{figure}

We compare NB-AIPW, FB-AIPW, and CSB-AIPW using the RD as the estimand. Table~\ref{tab:hct_result} summarizes the results obtained thus far, with the exception of the FRT $p$-value, which will be computed in Section~\ref{sec:frt-code}. Compared with FB-AIPW, CSB-AIPW produces an estimate closer to NB-AIPW while still improving precision through borrowing. Specifically, CSB selects 153 of the 335 matched ECs (see Figure~\ref{fig:sel}), yielding a smaller standard error and p-value than NB-AIPW, whereas FB-AIPW appears to overestimate the treatment effect by borrowing all matched ECs. Note that these results are based on \emph{synthetic data} generated to mimic the CALGB 9633 and NCDB studies and therefore differ from those reported in \cite{liu2025robust}; they are presented for illustrative purposes only.

Figure~\ref{fig:gamma} presents a sensitivity analysis over $\gamma$. The MSE-minimizing $\hat{\gamma}$ yields an ATE estimate close to NB with a relatively small standard error, illustrating the bias-variance trade-off in selecting $\gamma$.

\subsubsection{Penalized Selective Borrowing}

Next, we illustrate how to implement Penalized Selective Borrowing (PSB) using the \texttt{SelectiveIntegrative} package. We first create the required input objects using the package-specific function \texttt{SelectiveIntegrative::dataInput()}, and then fit the PSB model using \texttt{srEC()}. Here, we specify a generalized linear model for both the outcome model and the initial bias model by setting \texttt{method = "glm"}. Finally, we summarize the treatment effect estimates and the number of borrowed ECs.

\begin{rcode}
# data input
data_rct_psb <- SelectiveIntegrative::dataInput(
  dat_rct,
  y ~ treat * (sex + age + race + hist + tsize),
  treat ~ 1
)
data_ec_psb <- SelectiveIntegrative::dataInput(
  dat_ec,
  y ~ treat * (sex + age + race + hist + tsize),
  treat ~ 1
)
# run PSB
set.seed(1234)
result_psb <- srEC(
  data.rct = data_rct_psb,
  data.ec = data_ec_psb,
  method = "glm",
  rct.trControl = caret::trainControl(method = "cv", number = 10L),
  ec.trControl = caret::trainControl(method = "cv", number = 10L)
)
# summarize result
n_ec_total <- nrow(dat_ec)
n_ec_selected <- length(result_psb$subset.idx)
result_psb_summary <- tibble(
  method = c("NB-AIPW", "FB-ACW", "PSB-ACW"),
  tau.hat = c(
    result_psb$aipw$tau.hat,
    result_psb$acw$tau.hat,
    result_psb$acw.lasso$tau.hat
  ),
  sd.hat = c(
    result_psb$aipw$sd.hat, result_psb$acw$sd.hat, result_psb$acw.lasso$sd.hat
  ),
  ci.lower = tau.hat - qnorm(0.975) * sd.hat,
  ci.upper = tau.hat + qnorm(0.975) * sd.hat,
  p.value = 2 * pnorm(-abs(tau.hat / sd.hat)),
  n.sel = c(0, n_ec_total, n_ec_selected)
)
result_psb_summary
\end{rcode}

% \begin{verbatim}
% ## # A tibble: 3 x 4
% ##   method                         tau.hat sd.hat n.borrow
% ##   <chr>                            <dbl>  <dbl>    <dbl>
% ## 1 No Borrow AIPW                  0.0565 0.0457        0
% ## 2 Full Borrow ACW                 0.0761 0.0394      335
% ## 3 Penalized Selective Borrow ACW  0.0562 0.0427        1
% \end{verbatim}

\begin{table}[ht]
\centering
\caption{Estimation and asymptotic inference results from three borrowing strategies.}
\label{tab:psb_result}

\begin{threeparttable}
\setlength{\tabcolsep}{10pt}

\begin{tabular}{lccccc}
\toprule
Method & Estimate & SE & 95\% CI & Asym. $p$ & $n_{\rm borrow}$ \\
\midrule
NB-AIPW  & 0.0565 & 0.0457 & $(-0.0331,\;0.1461)$ & 0.2163 &   0 \\
FB-ACW   & 0.0761 & 0.0394 & $(-0.0011,\;0.1533)$ & 0.0534 & 335 \\
PSB-ACW  & 0.0562 & 0.0427 & $(-0.0275,\;0.1399)$ & 0.1881 &   1 \\
\bottomrule
\end{tabular}

\begin{tablenotes}[flushleft]
\footnotesize
\item NB = no borrowing; FB = full borrowing; PSB = penalized selective borrowing; AIPW = augmented inverse probability weighting; ACW = augmented calibration weighting.
\item Estimate = treatment effect estimate (\texttt{tau.hat}); SE = estimated standard error (\texttt{sd.hat}); 95\% CI = asymptotic 95\% confidence interval (\texttt{ci.lower} and \texttt{ci.upper}); Asym.\ $p$ = asymptotic $p$-value (\texttt{p.value}); $n_{\rm borrow}$ = number of borrowed external controls (\texttt{n.sel}).
\end{tablenotes}

\end{threeparttable}
\end{table}

Table~\ref{tab:psb_result} shows the results. PSB selects only one EC, producing an estimate nearly identical to the NB. This indicates that, when a GLM is used for binary outcomes, the penalized bias estimation procedure may lead to conservative borrowing decision in this dataset.

\subsection{Fisher Randomization Test}
\label{sec:frt-code}
Finally, we implement FRTs for NB, FB, and CSB. We use \texttt{ec\_borrow()} with the same arguments as in the previous analyses and additionally specify \texttt{n.fisher = 1000L} to perform 1,000 randomization permutations. To accelerate computation, we set \texttt{n.cores = 10} for parallel computing. Users should adjust this value according to their computing environment, which can be checked using \texttt{parallel::detectCores()}.

\begin{rcode}
result_nb_frt <- ec_borrow(
  data.rct = dat_obj[[1L]],
  data.ec = dat_obj[[2L]],
  outcome.type = "bin",
  method = "NbCovAdj",
  n.fisher = 1000L,
  n.cores = 10
)
result_fb_frt <- ec_borrow(
  data.rct = dat_obj[[1L]],
  data.ec = dat_obj[[2L]],
  outcome.type = "bin",
  method = "AIPW",
  n.fisher = 1000L,
  n.cores = 10
)
result_csb_frt <- ec_borrow(
  dat_obj[[1L]],
  dat_obj[[2L]],
  outcome.type = "bin",
  method = "CfAIPW",
  gamma.select = gamma_hat,
  seed = 1234,
  cf.regressor = Regressor("NN", list(k = 1)),
  cf.control = list(score = "NN", label.conditional = TRUE),
  n.fisher = 1000L,
  n.cores = 10
)
\end{rcode}

% # summarize result
% result_rd_frt <- bind_rows(
%   result_nb_frt,
%   result_fb_frt,
%   result_csb_frt
% ) %>%
%   filter(tau == "RD") %>%
%   select(-tau) %>%
%   mutate(
%     method = c("NB-AIPW", "FB-AIPW", "CSB-AIPW"),
%     .before = everything()
%   )

% \begin{verbatim}
% ## # A tibble: 3 x 9
% ##   method   tau.hat sd.hat ci.lower ci.upper p.value n.sel ess.sel FRT_p.value
% ##   <chr>      <dbl>  <dbl>    <dbl>    <dbl>   <dbl> <int>   <dbl>       <dbl>
% ## 1 NB AIPW   0.0568 0.0474  -0.0361    0.150 0.231       0      0        0.227
% ## 2 FB AIPW   0.0978 0.0376   0.0241    0.172 0.00932   335    331.       0.120
% ## 3 CSB AIPW  0.0584 0.0402  -0.0205    0.137 0.147     153    140.       0.187
% \end{verbatim}

Compared with the output presented in Section~\ref{sec:csb-code}, the FRT analysis additionally provides the FRT $p$-value reported in Table~\ref{tab:hct_result}.
In this example, both FB-AIPW and CSB-AIPW produce smaller FRT p-values than NB-AIPW, reflecting the efficiency gains obtained through borrowing ECs.

\section{Conclusion}
\label{sec:con}

Hybrid controlled trials provide a practical framework for improving the efficiency of RCTs by incorporating ECs. However, valid borrowing requires careful consideration of both covariate shift and outcome drift. In this tutorial, we presented a statistical analysis roadmap for HCTs, including eligibility alignment, matching, full borrowing, selective borrowing, and Fisher randomization tests. We also demonstrated how to implement these methods using the \texttt{SelectiveIntegrative} and \texttt{intFRT} packages through a reproducible workflow based on a synthetic lung cancer dataset included in the \texttt{intFRT} package. The examples illustrate that selective borrowing can improve efficiency while reducing the risk of bias from incompatible ECs, and that FRTs provide a complementary inference framework with finite-sample validity and post-selection validity.

Several limitations of this tutorial should be noted. First, the illustration uses a binary endpoint and does not address censoring, competing risks, or time-to-event endpoints. Second, the example focuses on analysis-stage borrowing after ECs have been assembled; design-stage assessment of EC fitness-for-purpose, data provenance, endpoint harmonization, and missing-data handling remains essential. Third, selective borrowing reduces sensitivity to outcome drift but cannot by itself validate an unsuitable EC source or remove bias due to unmeasured factors that are not reflected in the observed outcomes or covariates. Fourth, the tutorial emphasizes frequentist selective borrowing and randomization-based inference; Bayesian dynamic borrowing approaches are not discussed in detail.

Several important directions remain for future development. First, although this tutorial focuses on binary outcomes, many HCT applications involve \textbf{time-to-event endpoints} such as overall survival and progression-free survival. Extending selective borrowing and randomization-based inference in HCTs to survival outcomes remains an important area of ongoing research \citep{gao2025doubly}. Second, while this tutorial focuses on statistical analysis, developing \textbf{sample size determination} methods and design-stage tools for HCTs would facilitate their practical use in future clinical studies \citep{gao2025designing,liu2025sample}. Third, recent advances suggest that incorporating \textbf{super-covariates}, such as prognostic scores derived from machine learning models trained on external data sources, may further improve statistical efficiency while preserving validity \citep{schuler2022increasing,liao2025prognostic,hojbjerre2025powering}. Integrating such super-covariates into the proposed framework is a promising direction for future research.

Beyond methodological development, improving the implementation, accessibility, and interpretability of HCT methods is equally important. User-friendly software, transparent analysis workflows, and interpretable borrowing decisions can facilitate the routine application of HCTs in practice, improve communication among statisticians, clinicians, and regulators, and ultimately support broader regulatory acceptance of EC-assisted evidence.

\if1\anon
{
\vspace{-15pt}

\section*{Acknowledgment}
This project is supported by the Food and Drug Administration (FDA) of the U.S. Department of Health and Human Services (HHS) as part of a financial assistance award U01FD007934, \$2,556,429 over three years, funded by FDA/HHS. 
This work is also supported by R01AG066883, funded by the NIH/HHS.
The contents are those of the authors and do not necessarily represent the official views of, nor an endorsement by, the FDA/HHS, the National Institutes of Health, or the U.S. Government.
We used ChatGPT 5 to review grammar and improve the writing.

% \section*{Disclosure statement}\label{disclosure-statement}

% The authors have no conflicts of interest.

\vspace{-15pt}

\section*{Data Availability Statement}\label{data-availability-statement}

The synthetic dataset \texttt{lungcancer} used in this tutorial is included in the \texttt{intFRT} R package, which is publicly available at \url{https://github.com/IntegrativeStats/intFRT}.
} \fi

\vspace{-15pt}

\section*{Supplementary Material}
Supplementary materials include a complete R Markdown file with technical details, including package installation, runtime recording, and code for generating the plots.
They are also available at \url{https://github.com/ke-zhu/selective-borrow-tutorial}.

%Reproducible code, computational details, and supporting materials will be made available in an accompanying supplementary R Markdown file upon publication.

%\phantomsection\label{supplementary-material}
% \bigskip

% \newpage
\vspace{-15pt}
\bibliography{ref.bib}

% \newpage

\appendix

\end{document}